\documentclass[useAMS,usenatbib,usegraphicx]{mn2e}
\usepackage{amssymb}
\usepackage{graphicx}
\usepackage{subfigure}
\usepackage{longtable}

\newif\ifAMStwofonts

\def\kms{km s$^{-1}$}
\def\hi{H{\sc i}}
\def\hii{H{\sc ii}}
\def\msun{M$_{\odot}$}

\def\msunyr{M$_\odot$ yr$^{-1}$}
\def\cmdos{cm$^{-2}$}
\def\cmtres{cm$^{-3}$}
\def\x{$\times$}
\def\deg{$^{\circ}$}

\def\ldot{L$_\odot$}

\newcommand{\effm}{E_{\rm ff}}

\newcommand{\tham}{\theta_{\rm am}}
\newcommand{\lpcm}{l_{\rm pc}}
\newcommand{\dkpcm}{d_{\rm kpc}}
\newcommand{\emum}{\rm\,pc\,cm^{-6}}
\newcommand{\am}{$^{\prime}$}

\newcommand{\eff}{$E_{\rm ff}$}
\newcommand{\dkpc}{$d_{\rm kpc}$}
\newcommand{\emu}{$\rm\,pc\,cm^{-6}$}

\newcommand{\h}{$^h$}
\newcommand{\m}{$^m$}
\newcommand{\s}{$^s$}

\title[The ring nebula around HD\,211564]{The radio and IR counterparts of 
the ring nebula around HD\,211564}

\author[C. Cappa, J. Vasquez, S. Pineault, and S. Cichowolski]
{C.E. Cappa$^{1,2}$\thanks{Member of Carrera del Investigador, CONICET, 
Argentina\ E-mail: ccappa@fcaglp.unlp.edu.ar},
J. Vasquez$^{1,2}$\thanks{Postdoctoral fellow of CONICET, Argentina}, 
S. Pineault$^{1,3}$,
and S. Cichowolski$^{4}$\thanks{Member of Carrera del Investigador, CONICET, 
Argentina}\\
$^{1}$Instituto Argentino de Radioastronom\'{\i}a (CCT-La Plata, CONICET), 
CC 5, 1894, Villa Elisa, Argentina\\
$^{2}$Facultad de Ciencias Astron\'omicas y Geof\'{\i}sicas, Universidad 
Nacional de La Plata, La Plata, Argentina\\
$^{3}$D\'epartment de physique, de g\'enie physique et d'optique and Centre 
de recherche en astrophysique du Qu\'ebec (CRAQ),\\ Universit\'e Laval, 
Qu\'ebec, Canada GIVOA6\\
$^4$Instituto de Astronom\'{\i}a y F\'{\i}sica del Espacio (IAFE), CC 67,
Suc. 28, 1428 Buenos Aires, Argentina}

\date{Accepted 2005 June 23, Received 2005 June 22; in original form 2005 March 23}

\pagerange{\pageref{firstpage}--\pageref{lastpage}} 

\pubyear{2009}

\begin{document}

\maketitle

\label{firstpage}

\begin{abstract}

We report the detection of the radio and infrared counterparts of
the ring nebula around the WN3(h) star HD\,211564 (WR\,152), located to 
the southwest of the \hii\ region Sh2\,132. Using radio continuum data
from the Canadian Galactic Plane Survey, we identified the radio 
counterparts of the
two concentric rings, of about 9\arcmin\  and 16\arcmin\ in radius, 
related to the star. After applying a filling factor $f$ = 0.05-0.12, 
electron densities and ionized masses are in the range 10-16 \cmtres\ and 
450-700 \msun, respectively. 
The analysis of the \hi\ gas emission distribution allowed the identification
of \hbox{5900} \msun\ of neutral atomic gas with velocities between 
--52 and  --43 \kms\ probably linked to the nebula.
The region of the nebula is almost free of molecular gas. Only four small 
clumps were detected, with a total molecular mass of 790 \msun. About 310 \msun\ 
are related to a small infrared shell-like source linked to the inner ring, 
which is also detected in the MSX band A. An IRAS YSO candidate is detected in 
coincidence with the shell-like IR source.

We suggest that the optical nebula and its neutral counterparts 
originated from the stellar winds from the WR star and its massive 
progenitor, and are evolving in the envelope of a slowly expanding shell 
centered at $(l,b)$ = (102\fdg 30, --0\fdg 50), of about 31 pc in radius. 
The bubble's energy conversion efficiency is in agreement with recent numerical
analysis and with observational results.

\end{abstract}


\section{Introduction}\label{intro}

The strong mass flow and high UV photon flux of Wolf-Rayet stars originate
the so called {\it ring nebulae}, which can be identified as ring-like
optical features in the environs of the stars.
Searches for optical ring nebulae around  WR stars performed in the last 
decades (e.g. Chu 1981; Heckathorn et al. 1982; Miller \& Chu 1993; Marston et
al. 1994, 1995), 
resulted in the detection of more than 60 ring nebulae, 70\% of which 
are linked to WN stars. High angular resolution radio continuum observations 
led to the detection of the radio counterparts to 25\% 
of the ring nebulae, allowing the estimation of several physical
parameters (e.g. Cappa et al. 2002). 

In addition, spectral line observations of the neutral atomic 
and molecular gas helped to determine the characteristics of 
stellar wind bubbles, since they revealed large amounts of
neutral gas  linked to the nebulae, and allowed to include neutral 
gas in the estimate of the bubble's energetics (Cappa 2006).
Conclusions  from these studies, which indicated that a small amount of the
stellar wind energy released by the central star is converted into kinetic 
energy of the bubbles, are compatible with both observational and theoretical
studies (e.g. Chu et al. 1983; Oey 1996; Freyer et al. 2003, 2006; 
Cooper et al. 2004). 

In the last years, available infrared images obtained from the MSX Galactic
Plane Survey (Price et al. 2001) and GLIMPSE survey (IRAC images, Benjamin 
et al. 2003) allowed the identification
of interstellar bubbles in the near- and mid-IR, showing the presence
of photodissociation regions at the interface between the ionized and
molecular gas (Churchwell et al. 2006, Watson et al. 2008, Watson et al. 
2009).

\begin{figure*}
   \includegraphics[angle=0,width=160mm]{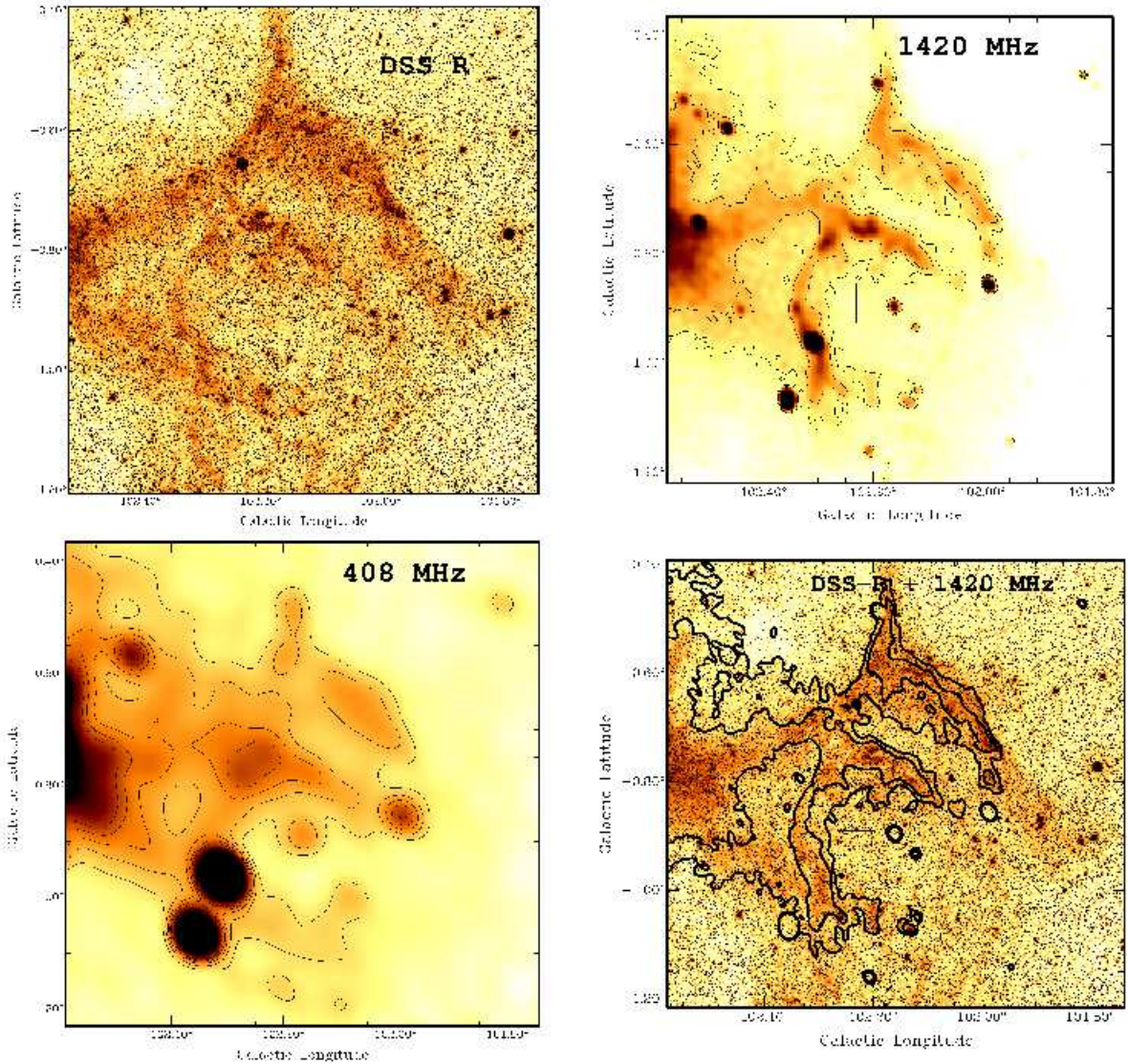}
  \caption{{\it Top left panel:} DSS2\,R image of the ring nebula around 
HD\,211564. The cross marks the position of the WR star. The color scale is 
in arbitrary units. 
{\it Top right panel}: CGPS radio continuum image at 1420 MHz. 
Contour lines are 7.0 and 7.4 K. 
{\it Bottom left panel}: CGPS radio continuum image at 408 MHz. 
Contour lines are 62, 65, 68, 71, 74, 100, 150, 
and 200 K. 
{\it Bottom right panel:} Overlay of the optical (in color scale) and 1420 
MHz (contour lines) images.}
\end{figure*}

Among the ring nebulae detected optically, the one around the WN star 
HD\,211564 is particularly interesting.  
The morphology of this optical ring nebula can be described by two
concentric ring structures having about 17\arcmin\ and 36\arcmin\ in diameter,
detected in [OIII], [SII], and H$\alpha$ + [NII] lines (Heckathorn et al.
1982).
The nebulae are easily identified in the DSS2\,R image 
displayed in the upper left panel of Fig. 1. The symmetry of the inner ring
is suggestive of a quite homogeneous interstellar medium, except
possibly in the south-southwest quadrant. 
Both structures are brighter in [OIII] lines than in H$\alpha$ + [NII], 
following the tendency for nebulae around early WN stars (Heckathorn et
al. 1982).
The inner structure is sharp in appearance and almost complete, 
with the WR star displaced 3\farcm 7 northeast from the geometrical
centre. On the contrary, the outer shell is more diffuse and less well
defined.

This ring nebula is located at the southwest of the \hii\ region 
Sh2-132,  whose brighter section (not shown here) is excited by the 
WR star HD\,211853, at $(l,b)$ = (102\fdg 78, \hbox{--0\fdg 65)},
and a few O-type stars.  
Although both HD\,211853 and HD\,211564 are linked to Sh2-132,   
they appear related to quite distinct structures.

Previous radio observations of Sh2-132 at 1420 MHz (Felli \& Churchwell
1972)
show a relatively large area of faint emission extending towards the 
region where the ring nebula is present.
The inner shell was also detected in the far infrared by Marston (1996),
who found an infrared shell of 21\arcmin\ in diameter 
surrounding the WR star.

HD\,211564 ($\equiv$ LS\,III+55\,30 = WR\,152) is a WN3(h) star 
(Smith et al. 1996)
located at $(l,b)$ = (102\fdg 23, --0\fdg 89) or
(RA,Dec[J2000]) = (22\h 16\m 24\s, +55\deg 37\arcmin 37\arcsec). 
The star is considered  a probable member of the Cep OB1 
association (Lundstron \& Stenholm 1984)
located in the Perseus spiral arm. 

The terminal wind velocity of HD\,211564 is in the range 1800-2100 \kms\
(Rochowicz \& Niedzielski 1995; Hamann \& Koesterke 1998; Niedzielski \& 
Skorzynski 2002),
while a mean value for the mass loss rate of WN-w
stars is about  5\x 10$^{-6}$\,\msunyr\  (Crowther 2007).

In this paper, we report on the radio counterpart of the ring nebula around
the WN star HD\,211564 first identified by Heckathorn et al. (1982)
and analyze the 
distribution of the ionized and neutral gas using radio continuum and \hi-21cm
line data from the  Canadian Galactic Plane Survey (CGPS, 
Taylor et al. 2003),
CO data from the Five College Radio Astrophysical Observatory
(FCRAO, Heyer et al. 1998), and infrared data from the IRAS and MSX satellites. 

\section{Data sets}

The analysis of the ionized and neutral atomic gas  in the 
environs of HD\,211564 was performed using data from the CGPS obtained 
with the Synthesis Telescope 
of the Dominion Radio Astrophysical Observatory (DRAO), in Canada.
This telescope performed interferometric observations of the 21-cm \hi\ 
spectral line, and, simultaneously, continuum emission in two bands centered 
at 1420 MHz and 408 MHz.  
Single-dish data were incorporated into the interferometric images ensuring
accurate representation of all structures to the largest scales.

To investigate the \hi\ emission distribution, we extracted a data cube 
centered at $(l,b,v)$ = (102\fdg 50, --1\fdg 0, --53.4 \kms)
and analyzed a region of 1\fdg 5\x 1\fdg 5 in size.
The synthesized beam is 1\farcm 19\x 0\farcm 98, the rms noise is
3 K in brightness temperature ($T_B$), and the velocity resolution and 
channel separation are 1.32  and 0.824 \kms, respectively.
The \hi\ images were convolved to a 2\farcm 5 \x 2\farcm 5 beam size to 
facilitate the identification of structures. The observed velocities cover 
the range --165 to +57  \kms.

The radio continuum data have synthesized beams of 3\farcm 4\x 2\farcm 8 
and 1\farcm 0\x 0\farcm 82 at 408 and 1420 MHz, respectively. The measured 
rms image noises are 1.1 and 0.063 K at 408 and 1420 MHz, respectively. 

The $^{12}$CO(1-0) line data at 115 GHz were obtained from Brunt \&
Heyer (2009). The angular resolution is 45\arcsec, 
the velocity resolution
1.0 \kms, and the rms noise, 0.7 K (main beam brightness temperature).

Infrared images at different wavelengths were used to investigate the dust 
distribution in the region of the nebula. High resolution {\it IRAS} 
images (HIRES) at 12, 25, 60, and 100 $\mu$m, were taken from IPAC
\footnote{IPAC is funded by NASA as part of the IRAS extended 
mission under contract to Jet Propulsion Laboratory (JPL)
and California Institute of Technology (Caltech).}. The angular resolution
is in the range 1\arcmin -2\arcmin.

Images at 8.28, 12.13, 14.65, and 21.3 $\mu$m (bands A, C, D and E, 
respectively) taken by the MSX satellite were also obtained from IPAC 
(18\farcs 4 in angular resolution). 

Additional radio and optical images of the region were retrieved from
the Skyview web page\footnote{http://skyview.gsfc.nasa.gov/}.

Finally, to investigate the presence of stellar formation activity
in the outskirts of the nebula, we used
infrared point sources from the MSX, 2MASS, and 
IRAS catalogues.

\section{Results}
\begin{figure}
    \includegraphics[width=7.3cm]{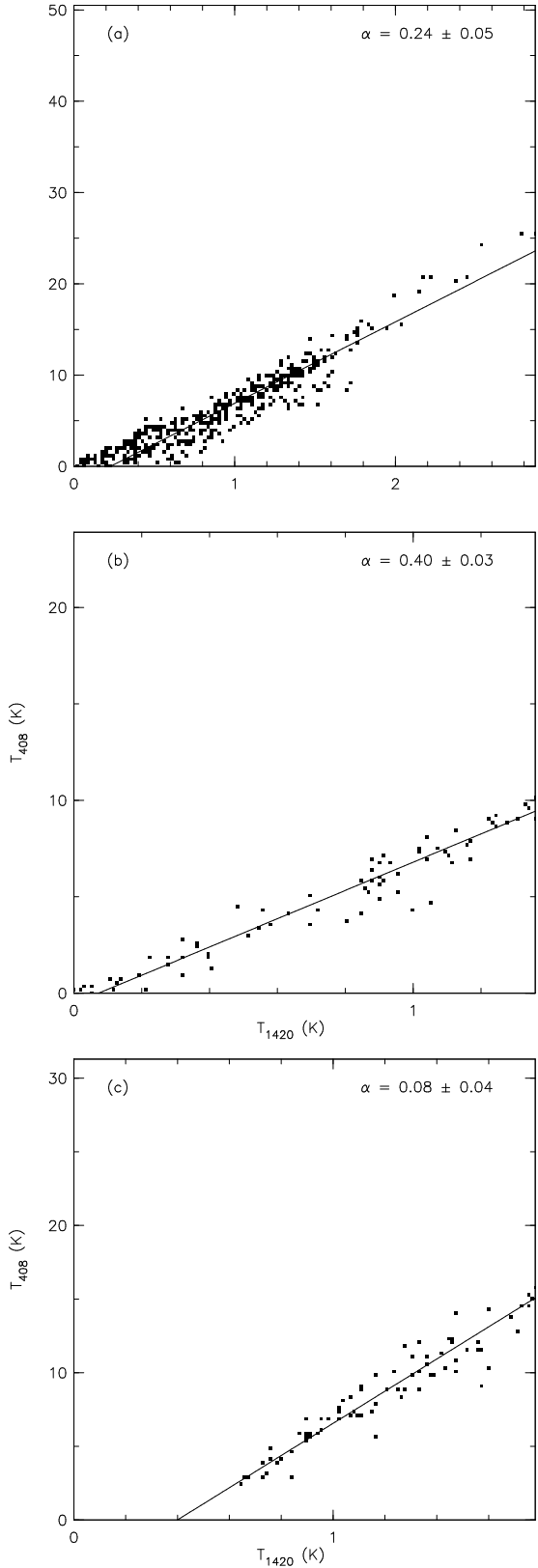}
\caption{TT-plots corresponding to (a) {\bf the entire region within an angular
radius of 24\arcmin\ from the center of the inner ring}, (b) the 
northwestern outer ring and
(c) the northern half of the inner ring.
For all plots, a background of 6 K and 57 K has been subtracted
from the convolved images at 1420 and 408 MHz, respectively.
}
  \label{figtt}
\end{figure}

\subsection{The ionized gas}

The radio emission at 1420 and 408 MHz, along with an 
overlay of the optical and  radio emissions at 1420 MHz, are
displayed in Fig. 1. Both the inner and outer features 
identified at optical wavelengths have well detected radio counterparts, as 
observed at 1420 MHz (upper right panel).
The lower right panel shows the striking correlation between the optical
and  1420 MHz radio continuum. 
The inner ring is not complete. However, very low level radio continuum
emission is present near $(l,b)$ = (102\fdg 07,--1\fdg 00). 
The outer ring is incomplete both at optical and radio wavelengths.
On the other hand, at 408 MHz only the inner
structure can be easily identified, while emission linked  to the outer 
optical feature is  discernible only near $(l,b)$ = (102\fdg 0, --0\fdg 67)
(Fig. 1, bottom left panel). The inner and outer 
rings, as estimated from the optical and radio images, are about 
19\arcmin\  and 32\arcmin\ in size, respectively. 
 
The nebula is also detected at 2.7 GHz (F$\ddot{u}$rst et al. 1990)
and at 4.85 GHz (Condon et al. 1994). Faint emission at 8.35 and 14.85 GHz 
(Langston et al. 2000)
most probably associated with the ring nebula can also be identified.   

In order to estimate flux densities of both the inner and outer structures
at 1420 MHz and to derive the radio continuum
spectral index, we analyzed first the nature of the small-diameter radio
sources projected onto the optical nebula. The coordinates of such sources
are listed in Table 1, along with the measured  flux densities at 408 and 1420 
MHz, the spectral index derived from the CGPS data, the catalogued flux 
densities at other wavelengths, and their identification. 

Sources 1,  2, 4, 5, and  6 appear projected onto the 
nebula. The location of sources  7 and 8 suggests that they are 
probably unconnected to the features we are interested in. 
Sources 1, 2, 5, 6, and 8 are clearly non-thermal in nature, probably 
extragalactic sources. 

After subtracting the contribution of the compact sources, and taking 
into account different background emissions, we estimated the flux density 
of the inner and outer features around HD\,211564 
from the image at 1420 MHz. Our estimates are $S_{\nu}$ = 0.6$\pm$0.1 Jy
and $S_{\nu}$ = 1.2$\pm$0.1 Jy for the inner and outer structures, respectively.

The CGPS images were used to evaluate the spectral index distribution of the 
radio continuum emission around HD\,211564.  As a first step, we removed 
the point sources from the region of interest and convolved the
1420 and 408 MHz images to the same circular resolution of 3\farcm 4.
We then constructed TT-plots, in which the brightness temperature $T_B$ at 
one frequency is plotted point-by-point against the brightness
temperature at the other frequency.  The temperature spectral index $\beta$,
where $T_B \propto \nu^{\beta}$, is directly related to the slope of the 
straight line fitted by regression, 
$\beta = \log(\rm slope)/\log(1420/408)$. The usual flux density spectral index
$\alpha$ ($S_{\nu} \propto \nu^{\alpha}$) is simply $\alpha = \beta - 2$.

\begin{table*}
{\small
\caption{Compact radio sources towards the ring nebula around WR\,152.}
\label{tabla2}
\begin{tabular}{cccccccccc}
\hline\hline
$No.$ & $(l,b)$  & S$_{1420}$ & S$_{408}$ & $\alpha$ & S$_{365}$ & S$_{610}$
& S$_{1420}$ & S$_{4850}$ & Identification \\
  & (\deg) &  mJy & mJy &  & mJy  &  mJy & mJy  & mJy   \\
  &   & (1)  & (1)  &  & (2) & (3) & (4) & (5) \\
\hline\hline
1  & 102.355,--1.065 & 165$\pm$5 & 409$\pm$22  & --0.7$\pm$0.2    & 677$\pm$78 & 245 &   &  52$\pm$8 &  87GB[BWE91]\,2216+5517  \\
2  & 102.308,--0.961 & 187$\pm$6 & 630$\pm$20  & --1.0$\pm$0.1 & 725$\pm$56 & 325 &   &  91$\pm$12&  87GB,221520.4+552034 \\
3  & 102.206,--0.160 & 10$\pm$1 & ---  &     &            &     & 9.7$\pm
$0.6  &   & NVSS\,J221720+552309 \\
4  & 102.122,--0.932 & 5.8$\pm$0.5 & ---  &     &            &     & 6.6$\pm$0.6  &    &NVSSJ\,221556+553135 \\
5  & 102.161,--0.896 & 17$\pm$1  &  35$\pm$4   &  --0.6$\pm$0.3   &            &     & 13.6$\pm$0.6 &    &NVSS\,J221601+553446 \\
6  & 101.986,--0.856 & 44$\pm$1  &  87$\pm$4  & --0.6$\pm$0.1 &            &     &    &         &  7C2213+5516 \\
7  & 102.519,--0.743 & 32$\pm$1 &  --- &     &            & 40  &  23.0$\pm$0.9&  & NVSS\,J221731+555424 \\
8  & 102.466,--0.619 & 30$\pm$1 &  66$\pm$7 &  --0.6$\pm$0.3   &            & 65  & 19.2$\pm$0.8 &  & NVSS\,J221632+560118 \\
\hline
\end{tabular}
}
References: (1) this paper, derived using CGPS data; (2) Douglas (1996),
Gregory (1991); (3) Harten et al. (1978); (4) Condon et al. (1998); 
(5) Becker (1991)
\end{table*}

\begin{figure}
   \includegraphics[angle=0,width=84mm]{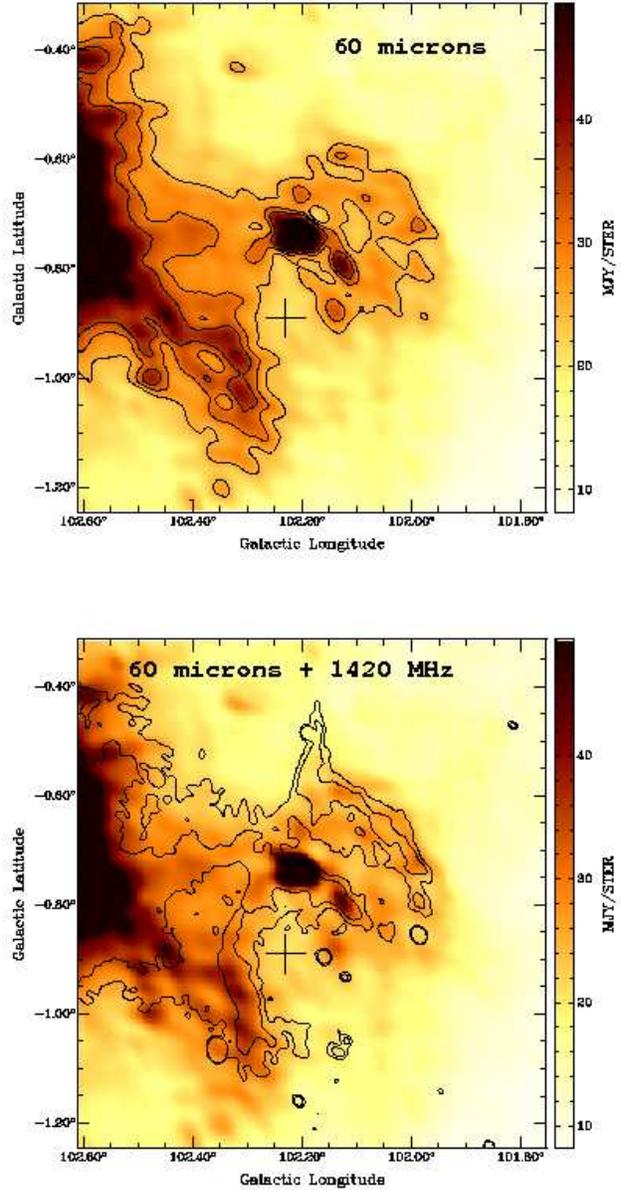}
  \caption{{\it Top panel:} IRAS (HIRES) image at 60 $\mu$m of the ring nebula
in color scale  and contours. 
Contours are 25, 30, 35, 50, 70, and 90 MJy ster$^{-1}$. 
{\it Bottom panel:} Overlay of the images at 60 $\mu$m (in color scale) and 
at 1420 MHz (in contours). }
  \label{fig1}
\end{figure}

The results of the TT-plot analysis are presented in {\bf Figure 2(a-c)}.
In each case, the analysis was limited to regions brighter than 6 K at 
1420 MHz and 57 K at 408 MHz. The first panel (Fig.~\ref{figtt}a) is a 
TT-plot of the entire region within an angular radius of 24\arcmin\ from 
the centre of the inner ring at $(l,b)$ = (102\fdg 20, --0\fdg 92).
Although a significant amount of scatter is present on the diagram, a 
convincing fit is obtained yielding $\alpha = 0.24 \pm 0.05$. Since this 
value is somewhat different from the canonical value of --0.1 for optically 
thin bremstrahlung, we
applied the same procedure to a number of sub-regions around HD\,211564 to look
for possible spatial variations.  Figures \ref{figtt}(b) and (c) show
the TT-plots corresponding to the northwestern outer ring near $(l,b)$ =
(102\fdg 08, --0\fdg 67) and the northern half of the inner ring,
respectively. There seems indeed to be significant variations.
As a further test, we evaluated the spectral index of the
nebulosities around nearby HD\,211853 and found $\alpha = 0.00 \pm 0.03$
(Vasquez et al. 2009),
a value clearly more consistent with optically thin bremstrahlung.

\begin{figure}
   \includegraphics[angle=0,width=80mm]{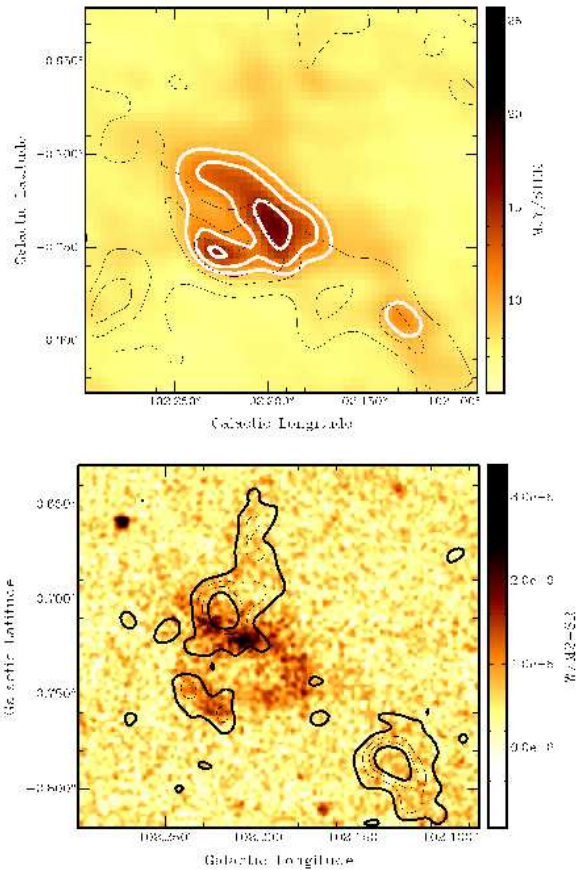}
  \caption{{\it Top panel:} Overlay of the emission at 25 $\mu$m (in
color scale and white contours) and the emission at 1420 MHz (in black 
contours). 
The white contours are 10, 12, and 15 MJy ster$^{-1}$,
while the black contours  are 7.4, 8.0, and 8.5 K.
{\it Bottom panel:} Overlay of the emission  at 8.3 $\mu$m 
(MSX band A) (color scale) and the CO(1-0) emission  (in
contours) within the velocity interval from  --49.4 to --46.4 \kms
(see Sect. 3.3). The contours are 3.0, 7.0, 11.0, and 15.0 K \kms. 
}
  \label{fig1}
\end{figure}

\subsection{Infrared emission}

The distribution of the infrared emission at 60 $\mu$m and its
correlation with the radio emission at 1420 MHz is displayed in Fig. 3.
The brighter sections of the inner shell are well detected in the far
infrared. Emission from the outer feature is hardly detected.
The far IR fluxes estimated for the ring nebula at 100 and 60  $\mu$m are 
$F_{100}$ = 660$\pm$130 Jy and $F_{60}$ = 340$\pm$100 Jy, respectively.
The quoted uncertainty was derived by using different values for the 
background emission.

Particularly interesting is the strong and extended IR source located
at $(l,b)$ = (102\fdg 20, --0\fdg 73). The source is
detected in the four IRAS bands. Its ring morphology is shown 
in the upper panel of Fig. 4, which displays an overlay of the emission at 
25 $\mu$m (color scale  and white contours) and in the radio continuum  at 
1420 MHz (black contours). 
This IR source partially coincides with a strong radio emitting region. 
The extended IR source has a clear counterpart in the MSX band A, although the 
emission is faint (Fig. 4, bottom panel). Note that the brightest region at
8.3 $\mu$m does not emit at 1420 MHz. Measured IR fluxes at 100, 60, 25, 
and 12 $\mu$m are 127$\pm$23, 53$\pm$3, 5.4$\pm$1.2, and 3.8$\pm$0.2 Jy, 
respectively. The estimated dust mass is 0.5$\pm$0.3 \msun.

Following the procedure described by Cichowolski et al. (2001), we derived
the color temperature of the dust  associated with the ring nebula 
and the strong IR source. 
Taking into account different values for the background emission, we 
found similar dust color temperatures $T_d$ =  30$\pm$4 K for the ring nebula
and the strong source.
The range of temperatures corresponds to  $n$ = 1-2, where the parameter 
$n$ is related to the dust absorption efficiency
($\kappa_\nu\ \propto\ \nu^n$). 
The dust temperature derived for both features is  typical for \hii\ regions.
The  dust mass linked to the nebula is 2$\pm$1 \msun.
 
\subsection{Neutral gas distribution}

\begin{table*}
\caption{IRAS point  sources classified as YSO candidates.}
\label{tabla3}
\begin{tabular}{cccccrrr}
\hline
 &  & & & \multicolumn{4}{c}{Fluxes} \\
No. & {\it (l,b)}  & Name  & 12$\mu$m  &    25$\mu$m &   60$\mu$m &  100$\mu$m & $L_{\rm FIR}$ \\
 & \deg &  & Jy & Jy & Jy & Jy & \ldot \\
\hline
1 & 102.122,--0.790  & 22135+5523 & 0.26 & 0.30  & 3.55   & 22.20 & 166\\
2 & 102.203,--0.723  & 22137+5529 & 0.91 & 0.80  & 20.75 & 74.75 &  705\\
3 & 102.440,--0.895  & 22158+5529 & 0.25 & 0.30  & 3.60   & 21.60 & 164 \\
\hline
\end{tabular}
\end{table*}

The analysis of the \hi\ datacube indicated that gas probably linked
to the ring nebula is present in the velocity interval from --57 to --37
\kms.
Figure 5 displays overlays of the  \hi\ emission within the velocity 
interval from  --55.5 to --39.0 \kms\ (in color scale) and the radio
continuum at 1420 MHz (contours). Each \hi\ image is the result of averaging 
the \hi\ emission within 3.3 \kms. 

The gallery of six \hi\ images was produced by removing the background 
in each channel map using DRAO software. The removed background  was estimated
as the average over the displayed image excluding a circular 
area of  28\arcmin\ in radius centered on the WR star. 
An inspection of the images shows that
the gas distribution is quite complex, showing several arc-like features.  

The \hi\ emission in the range --47 to --39 \kms\ (Fig. 5d, e, and f)
is characterized by the presence of a large cavity centered at 
$(l,b)$ = (102\fdg 30, 
--0\fdg 50), encircled by an almost circular envelope. The structure is
about 30\arcmin\ in radius. The inner and outer rings linked to the
WR star appear projected onto the weakest section of this large shell.
  
The images corresponding to the above mentioned velocity interval 
reveal the existence of neutral gas linked to the ring nebula. 
The presence of  \hi\ emission encircling the inner shell is easily 
identified between  --47 and --42 \kms\ near  $(l,b)$ = (102\fdg 35, 
--0\fdg 90) (Fig. 5d and e). This neutral gas forms a partial ring.
The outer ring appears 
also bordered by \hi\ gas near  $(l,b)$ = (102\fdg 00, --0\fdg 60). 
Note the presence of gas projected inside the inner ring. 

\hi\ emission in the velocity range from --52 to --49 \kms\ (Fig. 5b and c)  
is seen projected all over the area covered by the rings, being
even observed inside the inner ring and in the region between the
ionized rings. The partial \hi\ ring detected near $(l,b)$ = (102\fdg 35, 
--0\fdg 90) (Fig. 5d and e) is also detected in this velocity interval.
The emission extends towards $(l,b)$ = (102\fdg 08, --1\fdg 17), beyond 
the ionized arcs. Enhanced \hi\ emission bordering the inner ring is also 
present at --55.5 \kms\ (Fig. 5a).

The complex \hi\ gas distribution between --47 and --39 \kms\ suggests 
the existence  of two structures with similar velocity ranges partially
superpossed in the line of sight:  the large \hi\ shell
centered at  $(l,b)$ = (102\fdg 30, --0\fdg 50) and the partial \hi\ rings
linked to the ring nebula. 

The large \hi\ shell and its correlation with the optical emission is
shown in Fig. 6. The upper panel displays an overlay of the radio
continuum emission at 1420 MHz (in contours) and the \hi\ emission
between --47 and --37 \kms\ (in color scale). {\bf No background emission
was removed from the \hi\ image displayed in this fugure.} 
The middle panel displays
the DSS2-R image of the same area, and the bottom panel shows an overlay of 
the optical image and the \hi\ image of the upper panel (in contours). 
The saturated region at $(l,b)$ = (102\fdg 80, --0\fdg 70) in the 
optical image corresponds to the main body of the \hii\ region Sh2-132
(Vasquez et al. 2009). The  optical filament detected between
 $(l,b)$ = (102\fdg 40, --0\fdg 75) and $(l,b)$ = (102\fdg 17, --0\fdg 35)
and its faint extension towards $b >$ --0\fdg 35 defines an almost complete
ring-like structure coincident with the inner border of the large shell 
(see the bottom panel). Note that the brightest section of this filament
corresponds to the outer ring. No background was removed from the 
\hi\ image displayed in this figure.
\begin{figure*}
  \includegraphics[angle=0,width=175mm]{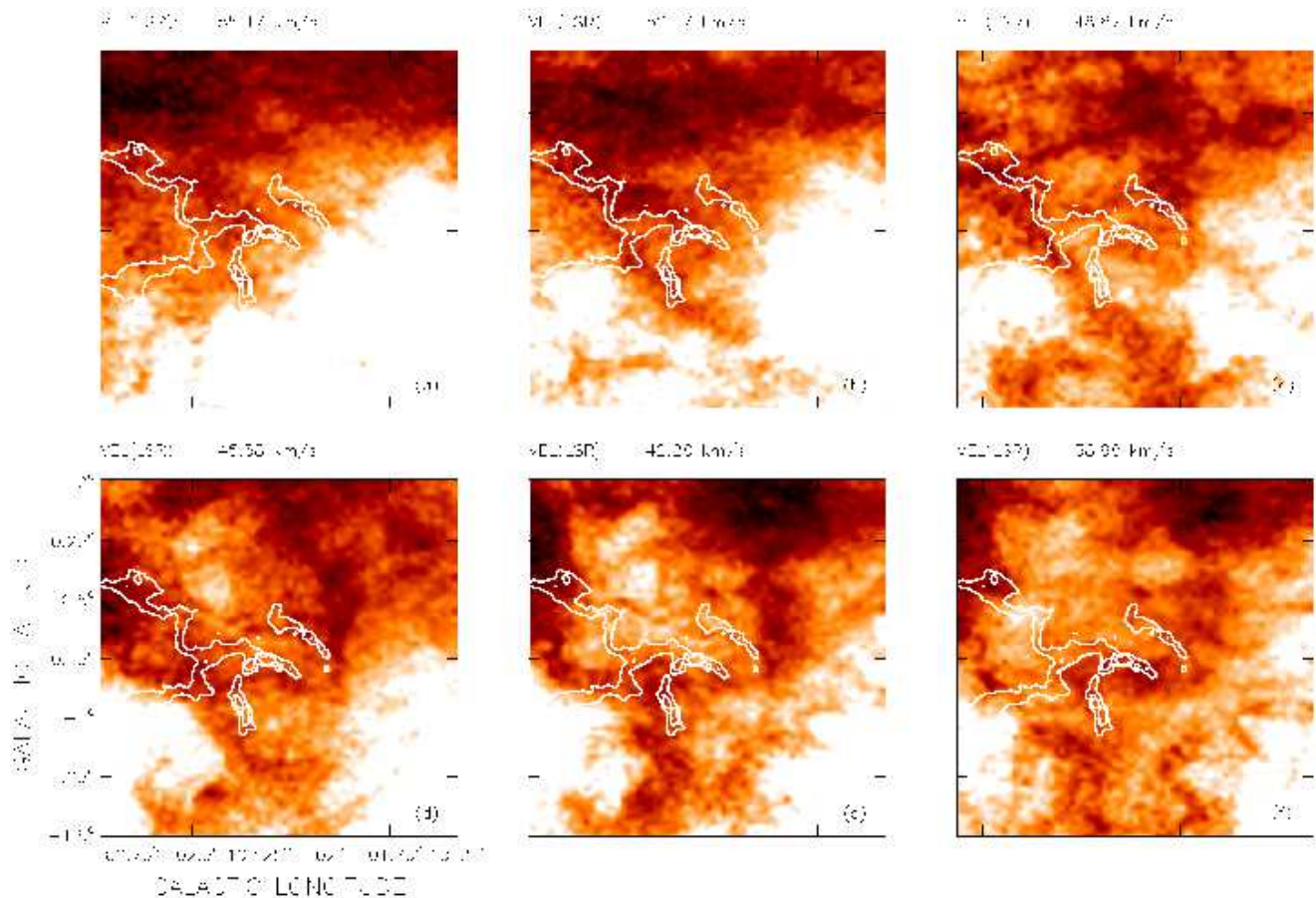}
  \caption{Radio continuum emission at 1420 MHz (contours) superimposed onto 
the averaged \hi\ brightness temperature in steps of 3.3 \kms\ between 
--55.5 and --39.0 \kms\ (color scale). The mean velocity  of each image is 
indicated in their upper left corner.
Contours are 7.4 and 8.0 K. Color scale goes from --15  to 40 K (see text).}
  \label{fig1}
\end{figure*}

To {\bf summarise}, gas clearly linked to the ring nebula is detected between --52 
and --43 \kms\ (Figs. 5b to 5e), while small patches of \hi\ gas are 
detected beyond these values. The large \hi\ shell centered at  
$(l,b)$ = (102\fdg 30, --0\fdg 50), of about 30\arcmin\ in radius, has an 
optical counterpart and is detected from --47 to --37 \kms\ in \hi\ emission. 
Neutral atomic gas belonging to these
structures is partially coincident in the line of sight.

\begin{figure}
   \includegraphics[angle=0,width=80mm]{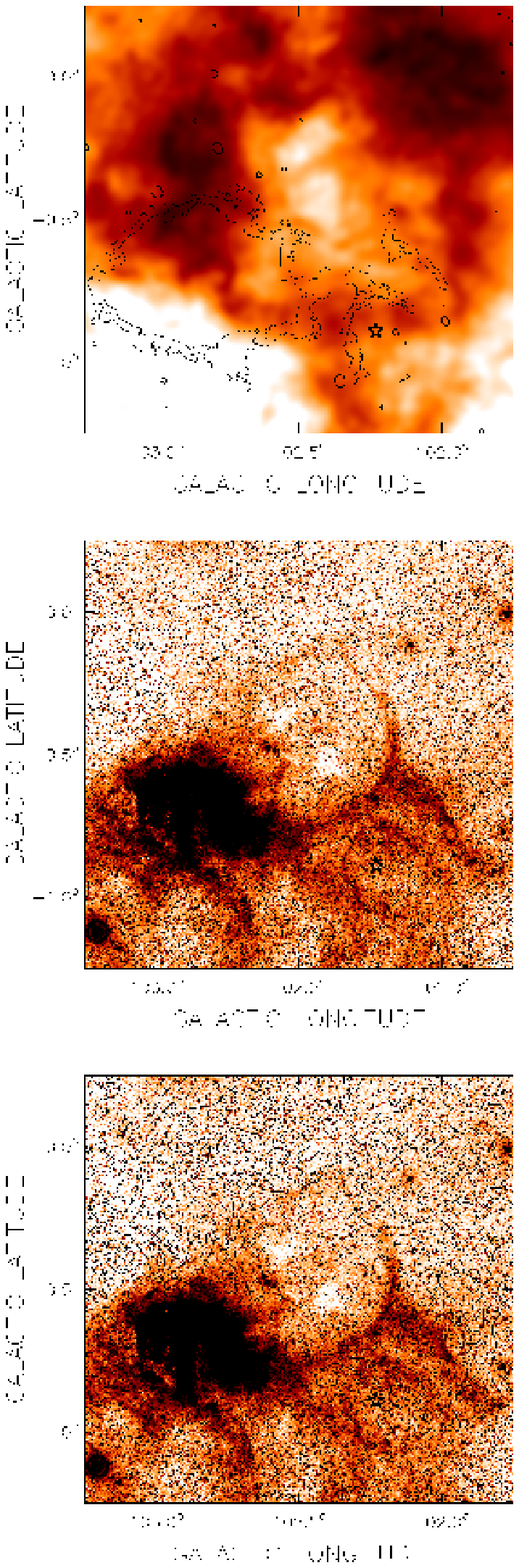}
  \caption{{\it Upper panel: } Overlay of the CGPS image at 1420 MHz 
(contours) and the \hi\ emission averaged within the velocity interval 
from --47 to --37 \kms. {\bf The color scale goes from 33 to 80 K. }
No background was removed from the \hi\ image.
Contours correspond to 7.4 and 8 K. 
{\it Middle panel:} DSS2-R image of the same
region. The intensity scale was chosen to show faint regions. 
{\it Bottom panel:} Superposition of the DSS2-R image and the \hi\ contours.
Contours are {\bf 48 to 72 K in steps of 4 K.}
}
  \label{fig1}
\end{figure}

\begin{figure}
   \includegraphics[angle=0,width=80mm]{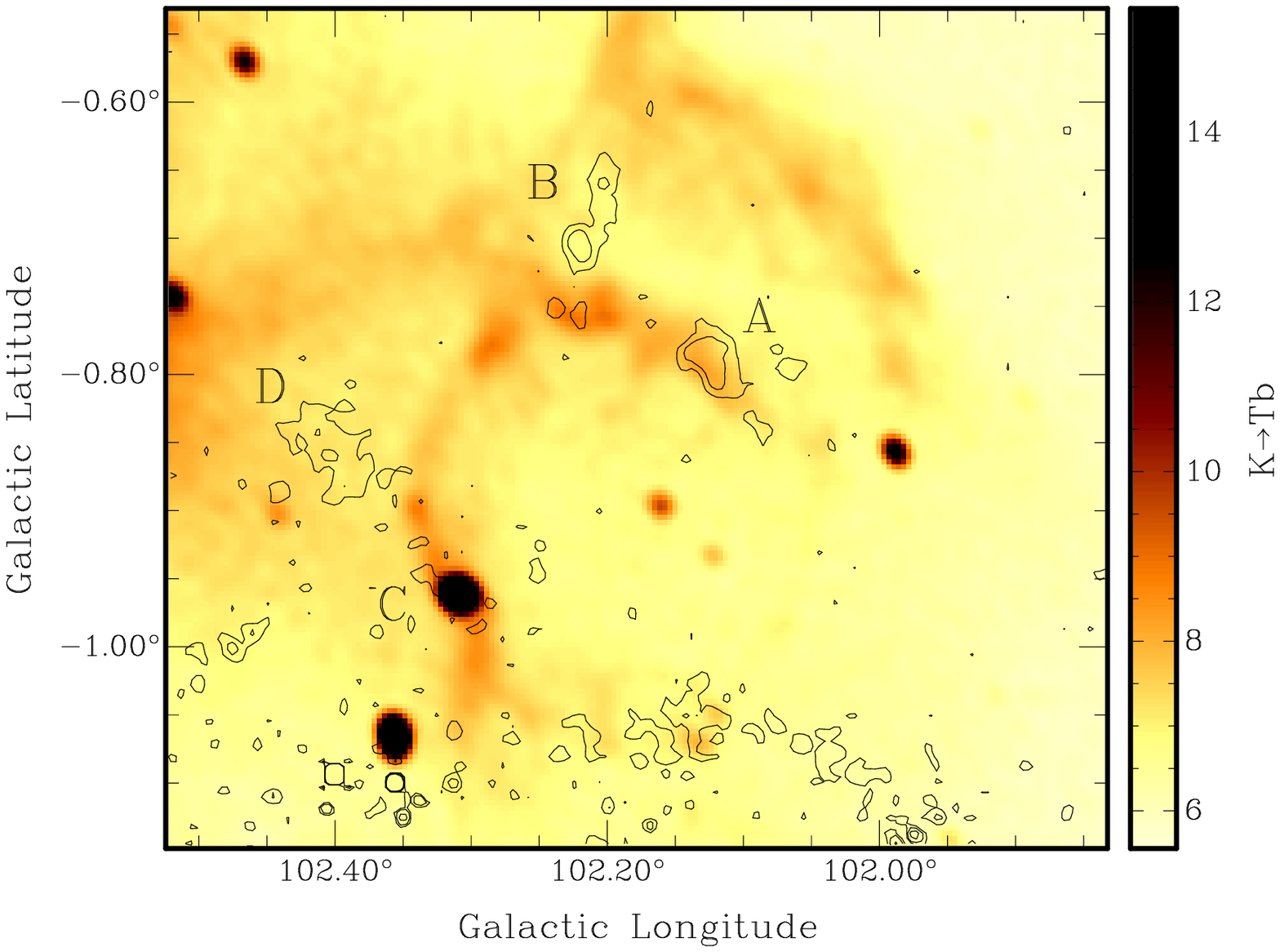}
  \caption{CO(1-0) integrated emission between --56.6 and --46.4 \kms\
(contours) superimposed onto the radio continuum emission {\bf at 1420 MHz} 
(color scale). 
Contours are 8.2, 14.3, and 20.4 K \kms.
}
  \label{fig1}
\end{figure}

The distribution of the molecular gas was investigated using the 
$^{12}$CO(1-0) datacube, in the velocity interval from --105 to +24 \kms. 
Within the area of 
interest, only a few patches of CO emission were  detected within the 
velocity ranges from --63 to --47 \kms, --30 to --28 \kms, and --9 to 
--2 \kms. Molecular gas having velocities $v <$ --47 \kms\
is probably linked to the ring nebula. The integrated emission
between --56.5 and --46.4 \kms\ is displayed in Fig. 7 superimposed
onto the radio continuum  emission. Only four molecular clumps are detected,
and three of them coincide with the optical rings
(clump A at  $[l,b]$ = [102\fdg 13, --0\fdg 80], clump B at $[l,b]$ = 
[102\fdg 23, --0\fdg 70], and clump C at $[l,b]$ = [102\fdg 32, 
--0\fdg 96]). 
Note also that CO clump D at $(l,b)$ = (102\fdg 38, --0\fdg 85)
{\bf and Clump B  
are projected onto regions} of low radio continuum emission.

CO near clump B is probably associated with the IR extended source 
described in Sect. 3.2, as is suggested by the striking spatial correlation
with the IR ring structure detected in the far and mid IR (see the
bottom panel of Fig. 4).
Note that the emission in  MSX band A is enhanced towards the borders
of Clump B at $(l,b)$ = (102\fdg 21, --0\fdg 73). 
The presence of molecular emission in the region indicates that the emission 
in band A most probably originates in  polycyclic aromatic 
hydrocarbons (PAHs), which can be found in photodissociation regions at 
the interface between ionized and molecular gas (e.g. Churchwell et al. 2006). 

We searched for young stellar object (YSO) candidates in the IRAS, MSX, and 
2MASS point source catalogues (Egan et al. 2003, Cutri et al. 2003).
Following criteria by Junkes et al. (1992), only three IRAS point sources 
have infrared colors compatible with an YSO classification.
Their galactic coordinates are listed in Table 2, along with their 
identifications, fluxes in the 
four IRAS bands, and IR luminosities, estimated following Yamaguchi et al. 
(1999). 
  
It is worth mentioning that source 2
coincides with the extended IR source described 
above and with clump B.
Source 1 appears projected onto clump A, while  source 3 might be
linked to Clump D. 
The presence of YSO candidates coincident with molecular clumps
suggests that star formation is  active in certain areas.




\begin{table}
\caption{$H_2$ column densities and masses.}
\label{tabla3}
\begin{tabular}{crrrr}
\hline
Cloud      & $W_{CO}$ &  $N_{H2}$  & Area  & $H_2$ mass \\
           & K \kms   &  10$^{20}$ \cmdos    & pc$^2$& \msun \\
\hline
A          & 9.9    & 10.5$\pm$1.4  & 14.2 & 310 \\
B          & 9.3    &  9.9$\pm$1.2  & 11.1 & 200 \\
C          & 6.6    &  7.0$\pm$0.9  &  2.7 &  40 \\
D          & 8.9    &  9.4$\pm$1.2  & 12.0 & 235 \\
\hline
\end{tabular}
\end{table}

\section{Discussion}

\subsection{The distance  of the ring nebula and the large \hi\ shell}

Since most of the physical parameters of the nebula depend on its 
distance, it is important to estimate its value.
Based on the association of the WR star to Cep OB1, and adopting $v$ = 
11.67 mag, and $(b-v)$ = 0.17 mag, van der Hucht (2001) 
places the star at a distance of 2.75 kpc, with an uncertainty of about 20\%. 
This distance implies an absolute magnitude $M_v$ = --2.15 mag for the 
WR star, somewhat different from the adopted mean value for the subclass 
(in the range --3.9 to --2.8 mag, see table 24 by van der Hucht [2001]).
A quite 
different spectrophotometric distance $d_*$ = 4.2 kpc can be estimated by 
adopting a mean value $M_v$ = 3.09 mag for WN3 stars  and the same 
absorption $A_v$ = 1.6 mag (van der Hucht 2001).

As shown in Sect. 3.3, \hi\ gas related to the ring nebula around HD\,211564
is present in the velocity range {\bf from \hbox{--52}} to --43 \kms, with a  mean 
velocity of $\approx$--47 \kms.
Molecular gas probably related to the inner structure has velocities
in the range from --49.4 to --46.4 \kms. Thus, a systemic velocity 
$v_{sys}$ = --47$\pm$2 \kms\ for the neutral features related to the star
seems appropriate.

The analytical fit to the circular galactic rotation model taking into
account non-circular motions (Brand \& Blitz 1993) predicts that gas 
having velocities of --47 \kms\ should be placed at a kinematical distance
of 3.5 kpc. 
This value is, within errors,  compatible with the derived stellar distance.
As a consequence, we adopt  3.5$\pm$1.0 kpc as the distance to the
WR star and the surrounding structure.
At this distance, the inner and outer ionized rings have 
linear radii of 8.3$\pm$2.4 pc and 15.4$\pm$4.4 pc, respectively. 

\hi\ gas belonging to the large shell has velocities in the range 
--46 to  --38 \kms, rather similar to the velocity of the gas linked to the 
ring nebula, suggesting a similar distance. As a working hypothesis,
we adopt a distance of 3.5$\pm$1.0 kpc for the large shell. At this 
distance, this shell is 31 pc in radius.  

\subsection{Scenario}

Given the observed morphology, a possible scenario to explain the 
presence of two concentric ionized structures is that the inner and 
outer rings have been formed by the action of the stellar winds from 
different evolutionary phases of the  star, as proposed by Garc\'{\i}a-Segura 
\& Mac Low (1995) for NGC\,6888. Such a
scheme has proved to explain the gas distribution in a number
of stellar wind bubbles showing double structures (e.g. the ring nebulae
around WR\,16 and WR\,85, Marston 1995).

However, the presence of neutral atomic gas  observed between the inner 
and outer rings, and bordering the inner ring, seems to 
conspire against such an interpretation for the case of HD\,211564.   
An alternative explanation is that both rings form only one structure, 
with its main axis nearly along the line of sight. 
This scenario is compatible with the detection of gas seen in projection
between both rings. 

Here, we propose that the stellar wind bubble associated with HD\,211564 
is evolving in the neutral gas envelope of the large shell centered at 
$(l,b)$ = (102\fdg 30, --0\fdg 50). 
Support to this interpretation comes from the fact that the outer ring 
near $(l,b)$ = (102\fdg 30, --0\fdg 70) appears arched, suggesting that 
it is being pushed from larger galactic longitudes and higher galactic 
latitudes. 
In this context, the section of the outer ring near (102\fdg 25, --0\fdg 60)
would be the result of the interaction of both structures. The fact that the
ring nebula is detected towards the fainter region of the large \hi\ shell
and the presence of \hi\ gas bordering the outer ring at $(l,b)$ =
(102\fdg 00, --0\fdg 60) reinforces the interpretation  that the interstellar
bubble is evolving in the compressed envelope. In this scenario, part of
the \hi\ gas seen inside the inner ring and between the outer and inner
rings might be linked to the large shell. The fact that the large shell
and the interstellar bubble have similar velocities makes it difficult to
clearly identify \hi\ gas belonging to each individual structure. 

Note that the ionized rings are detected where gas of the \hi\ envelope
centered at (102\fdg 35, --0\fdg 50) is present, i.e. where the ambient 
density is relatively high, in agreement with previous findings by 
Naz\'e et al. (2001) for bubbles in the LMC.
 
\subsection{Physical parameters of the nebula}
 
We now estimate the main physical parameters of the bubble associated
with HD\,211564 bearing in mind
the proposed scenario. Thus, we assume that the outer and inner rings form
only one structure with the  radius of the outer ring, i.e. 16.3$\pm$4.6 
pc.

The parameters of the ionized gas were derived from the image at 1420 MHz. 
Electron densities and \hii\ masses were 
obtained from the expressions by Mezger \& Henderson (1967) 
for a spherical 
\hii\ region of constant electron density (rms electron density $n_e$).
The presence of He\,{\sc ii} was accounted for by multiplying the 
\hii\ mass by 1.27. The rms electron density and ionized mass 
are 3.5 \cmtres\ and 2000 \msun,  respectively. Errors of 30\% and
60\% in the electron density and in the ionized mass come from
the distance uncertainty.

The distribution of ionized gas, as shown both by the optical and 
radio emissions, suggests filling factors in the range $f$ = 0.05-0.12
(estimated assuming a spherical bubble of outer radius equal to 16.3 pc,
and that 10 to 25\% of the surface is covered by plasma). 
By applying these factors, electron densities and ionized masses are 
10-16 \cmtres\ and 450-700 \msun, respectively. 

The number of UV photons necessary to ionize the gas in the inner and 
outer rings, as derived from radio continuum emission, is $\log \ N_{Ly-c}$ 
= 48.2, lower than the UV photon flux emitted by the WR star ($\log \ N_* $ 
= 49.2, Crowther 2007).
We can conclude that the WR star can maintain
the ionization of the rings. The difference between the two values is 
consistent 
with the fact that a large number of UV photons may escape from the bubble 
through the patchy neutral envelope and/or are 
absorbed by dust grains mixed with the ionized gas.

An important point to discuss in connection with the radio continuum emission
is the fact that the derived spectral indices seem to deviate from 
typical values for thermal emission.
If the large values of spectral index found for
some regions (e.g., Fig.~\ref{figtt}b) are genuine, 
this could imply that there are regions
where the thermal emission is approaching the optically thick regime.
The relative faintness of the emission however militates against such
an interpretation, unless the radio continuum emission originates from
regions considerably smaller than the observing beam.

If the emitting regions are in the form of small
clumps of typical angular size $\theta$,
or of thin filaments of angular thickness 
$\theta$, beam smearing or dilution will cause the emitted regions to
have an observed emission measure $EM_o$ smaller than the true
emission measure $EM$ by a factor $f_e$ where $f_e \approx \Omega_b/\theta^2$
for clumps and $f_e \approx \theta_b/\theta$ for filaments, and
$\Omega_b$ and $\theta_b$ are the beam angular area and beam angular
diameter, respectively. The resolution at 1420 MHz is about 1\arcmin\ so
that, writing $\tham$ as the typical angular size in arcmin, we simply
have $f_e \approx 1/\tham^2$ and $1/\tham$ for clumps and filaments,
respectively.

The brightness temperature due to optically thin bremstrahlung or free-free
emission $T_{\rm ff}$ is given by $T_{\rm ff} = T_e\,\tau_{\rm ff}$ where
$T_e$ and $\tau_{\rm ff}$ are the electron temperature and free-free optical
depth, the latter given by (e.g., Mezger \& Henderson
1967, Chaisson 1976):
\begin{equation}
\tau_{\rm ff} = 6.5 \times 10^{17}\,a(\nu,T_e)\,T_e^{-1.35}\,
     \nu^{-2.1}\, \effm,
\end{equation} 
where $a(\nu,T_e)$ is the Gaunt factor (about unity for our purposes) and
the emission measure \eff\ is in \emu.  Setting $\tau_{\rm ff} = 1$
in the above equation, and taking
$a=1$ and $T_e \approx 7000 \rm\, K$, the relation between the turnover
frequency (in GHz) $\nu_{\rm GHz}$ and free-free emission measure is given
by $EM = 1.8 \times 10^6\,\nu_{\rm GHz}^{2.1}$.  A turnover at 1 GHz
then corresponds to an emission measure of $1.8 \times 10^6 \emum$.

The observed emission measure is given in terms of the free-free brightness
temperature by $ EM_o \approx 570\,T_{\rm ff}$ which, 
for $T_{\rm ff} \approx 8 \rm\,K$,
gives $4500 \emum$ for $a=1$ and $T_e \approx 7000 \rm\, K$.  The required
beam dilution factor $f_e$ is thus of the order of 400.
To relate this factor to the typical size of clumps or thickness of filaments,
we note that the linear size of a structure in pc is related to its angular
size $\tham$ in arcmin by the relation $\lpcm = 0.29 \, \tham\,\dkpcm$, where
\dkpc\ is the source distance in kpc. For the assumed distance of 3.5 kpc,
this relation reduces simply to $\lpcm \approx \tham$.  We thus obtain $\lpcm \approx
1/(400)^{1/2}$ = 0.05 for clumps and $\lpcm \approx 1/400 = 0.0025$ for 
filaments.

To estimate the  neutral atomic mass we took into account the
\hi\ gas with velocities between --52 to --43 \kms\ projected onto 
the region of the nebula. This velocity interval corresponds to the
range in which \hi\ gas linked to the nebula can be more clearly
identified.
We assumed that the gas is optically thin and included a He abundance of 10\%.
The derived neutral gas mass is 5900 \msun. Uncertainties are about 70\%.

The $H_2$ column density and the molecular mass of the cloudlets  were 
estimated from the $^{12}$CO data, making use of the empirical relation 
between the integrated emission $W_{CO}$ (= $\int T dv)$ and $N_{H2}$. We 
adopted $N_{H2}$ = (1.06$\pm$0.14) \x W$_{CO}$ $\times$ 10$^{20}$ cm$^{-2}$ 
(K \kms)$^{-1}$, obtained from $\gamma$-ray studies of molecular clouds in 
the Orion region (Digel et al. 1995). The integrated emission, $H_2$ column 
density, area, and the molecular mass of each cloudlet are listed in 
Table 3. 

The small amount of molecular gas in the whole region suggests that
either little molecular material was present in the region when the 
massive star formed or most of the molecular gas was photodissociated
and ionized by the strong UV stellar flux. The presence of these small
molecular clouds is compatible with the poor signs of stellar formation  
in the region.

The ambient density derived by distributing the ionized, neutral atomic and 
molecular mass (5900 \msun) over a sphere of 16.3 pc in radius is  
$\simeq$14 \cmtres.
This value, higher than that of the typical interstellar medium, is 
compatible with the suggestion that the interstellar bubble evolved in 
a neutral expanding shell formed by compressed gas.
 
To verify whether HD\,211564 can provide the energy to blow
the interstellar bubble, we estimated the mechanical energy 
$E_\mathrm{w}$ released by the massive star into the ISM 
during the dynamical age $t_\mathrm{d}$ of the bubble
and compared it to the kinetic energy  $E_\mathrm{k}$ 
of the structure.

The kinetic energy of the interstellar bubble $E_\mathrm{k}$ 
= $M_\mathrm{b}v_\mathrm{exp}^2/2$ was derived adopting an expansion 
velocity of 9 \kms\ (based on the velocity range of the associated \hi\ 
gas), and assuming that the total mass related to the bubble 
$M_\mathrm{b}$ (ionized, neutral atomic, and molecular) amounts to 5900 
\msun. The kinetic energy is 4.8$\times$10$^{48}$ erg.

The dynamical age of a wind blown bubble can be estimated as 
$t_\mathrm{d}$ = 0.56$\times$10$^6 R/v_\mathrm{exp}$ yr 
(McCray 1983), where $R$ is the radius of the bubble (pc), 
$v_\mathrm{exp}$ is the expansion velocity (\kms), and the coefficient
is the deceleration parameter. Adopting $R$ = 16.3 pc and $v_\mathrm{exp}$
= 9 \kms, $t_\mathrm{d}$ = 1$\times$10$^6$ yr. Large uncertainties are 
involved in this result. 

The stellar wind mechanical energy $E_\mathrm{w} (= L_\mathrm{w}t 
= \dot MV_\mathrm{w}^2t/2$) released by the WR star can be estimated
from the stellar wind parameters. Since we cannot discard that the wind 
of the O-type star progenitor of the present WR star has contributed to 
the formation of the bubble, we take into account the contribution
of the star during the WR and main sequence phases.
The contribution during the WR phase can be derived adopting the stellar 
wind parameters listed in Sect. 1 
($\dot{M} = $ 5$\times$10$^{-6}$\,M$_\odot$\,yr$^{-1}$ and
$V_\mathrm{w}$ = 2000~\kms), and assuming that this stellar phase lasts
0.5$\times$10$^6$ yr, compatible with the lifetime of the WR 
phase of  a massive star (Meynet \& Maeder 2005). 
$E_\mathrm{w}$ results to be  1.0$\times$10$^{50}$ erg.
To estimate the contribution during the main sequence phase, we adopt 
mean stellar wind parameters for O-type stars, i.e. 
$\dot{M}$ = 2$\times$10$^{-6}$\,M$_\odot$\,yr$^{-1}$, $V_\mathrm{w}$
= 2000~\kms\ (Prinja et al. 1990, {\bf Mokiem et al. 2007)}, and 
assume that 
the stellar wind acted during at {\bf least 3$\times$10$^{6}$ yr.} The wind 
mechanical energy released during the main sequence phase of the star is 
then {\bf 1.4$\times$10$^{50}$ erg.}  Consequently, we adopt 
{\bf 2.4$\times$10$^{50}$} 
erg as the total energy released through stellar winds.
 
The energy conversion efficiency  $\epsilon$ 
(= $E_\mathrm{k}$/$E_\mathrm{w}$) = {\bf 0.02} is 
compatible with a stellar wind origin for the ring nebula. Uncertainties in
this estimate originate from the input stellar wind energy  (i.e. the
mass loss rates and terminal velocities), the adopted age of the bubble 
(which depends on the expansion velocity and the expansion law of the
bubble), and the kinetic energy (which includes masses and expansion 
velocity). 

The derived values are similar to estimates for other stellar
wind bubbles (e.g., Cappa 2006). They are in agreement with 
predictions from recent numerical
simulations from Freyer et al. (2003, 2006). These authors take into account  
the action of the stellar wind and the ionizing flux from stars of
35 and 60 \msun\ on the surrounding gas as they evolve from the main sequence 
stage to the presupernova phase, and investigate the ``missing wind 
problem''.
From their simulations, they find $E_\mathrm{k}$/$E_\mathrm{w} \approx$
0.10-0.04 for stars with 35 and 60 \msun, respectively. Among the additional 
solutions for the ``missing wind problem'' in superbubbles listed by 
Butt \& Bykov (2008), the presence of a blowout, in which the hot gas and
the energy can spew out from the bubble (Cooper et al. 2004) can clearly
apply to the present case.

\subsection{The large \hi\ shell}

Assuming a distance of 3.5 kpc for the large shell, the associated 
neutral atomic mass amounts to about 10$^{4}$ \msun. 
The analysis of the CO datacube reveals that a small amount of molecular
gas may be associated with the shell, indicating that the large shell
evolved in a region with an ambient density of about 3 \cmtres.

The kinetic energy  and the dynamical age estimated adopting an expansion 
velocity of 6-9 \kms\ is (4-8)$\times$10$^{48}$ erg and 
(1.9-2.6)$\times$10$^{6}$ yr, respectively. 

As regards the origin of the \hi\ shell, we note that, for an
energy conversion efficiency  of 0.10, an O-type star 
with stellar wind parameters  $\dot{M}$ = 
2$\times$10$^{-6}$\,M$_\odot$\,yr$^{-1}$ and  $V_\mathrm{w}$ = 1500~\kms\
might provide the energy to create the observed structure. However, no 
O-type star has been detected towards the inner part of the large shell.

\section{Summary}

In this paper we have analyzed the distribution of the ionized and neutral gas, 
as well as the dust particles, in the surroundings of the WR star HD\,211564,
located at about 3.5 kpc.

The  radio continuum data shows that the ring nebula related to 
HD\,211564 has a clear radio counterpart. Two concentric rings of about 
9\am\, and 16\am\, in radius are well identified at 1420 MHz. 
The spectral index of the radio continuum emission has been estimated in 
several regions of the nebula using the TT-plot method. We have found 
that the spectral index shows significant variations and that it slightly
deviates from the typical value  expected for \hii\ regions,
suggesting the presence of small and dense clumps of ionized gas
which are optically thick at 1420 MHz. 
The analysis of the available radio continuum data has enable us to 
estimate some parameters that characterize the \hii\ region. Considering 
a filling factor in the range 0.05-0.12, we obtained  ionized masses of 
450-700 \msun, {\bf and} electron densities of 10-16 \cmtres.

The analysis of the infrared emission shows that the brighter sections of 
the inner ring {\bf are} well detected in the far infrared, while the outer ring 
is hardly identified. A strong IR source, detected in the four IRAS bands, 
is observed at ($l,b$) = (102\fdg 20, --0\fdg 73). This source presents a 
ring morphology and is also detected in the MSX Band A.

An inspection of the \hi\ images suggests that the cavity blown by HD\,211564 
is evolving in the border of an expanding shell of about 30\arcmin\ in
radius, centred at  $(l,b)$ = (102\fdg 30, --0\fdg 50). This large structure 
is observed in the velocity interval from --47 to --39 \kms. 
On the other hand, \hi\ emission probably related to the ring nebula is 
present in the velocity range from --52 to --43 \kms. The fact that both 
structures are partially observed at the same velocities, together with 
their observed morphologies, suggest that both structures are interacting.
The derived neutral gas mass related to the ring nebula is 5900 \msun.

Molecular gas probably linked to the ring nebula is observed in four small 
clouds. 
We have found three YSO candidates {\bf associated with} molecular clouds. 
An energetic analysis shows that HD\,211564 emits enough ionizing photons 
to keep the gas ionized and heat the dust, and that its stellar wind is 
sufficient to explain the formation of the observed features.

\section{Acknowledgements}

We thank C. Brunt and M.H. Heyer for making their CO data
available in advance of publication. Provision of the CO data was 
supported by NSF grant AST0838222.
We acknowledge the anonymous referee for her/his comments. 
This project was partially financed by the Consejo Nacional de 
Investigaciones Cient\'{\i}ficas y T\'ecnicas (CONICET) of Argentina under 
projects PIP 112-200801-02488 and PIP 112-200801-01299,  Universidad Nacional 
de La Plata (UNLP) under project 11/G093, Universidad de Buenos Aires under 
project UBACyT X482, and Agencia Nacional de Promoci\'on Cient\'{\i}fica y 
Tecnol\'ogica (ANPCYT) under projects PICT 00812 and 2007-00902.
The Digitized Sky Survey (DSS) was produced at the Space Telescope Science
Institute under US Government grant NAGW-2166. This work was partly (S.P.) supported by the Natural Sciences
and Engineering Research Council of Canada (NSERC) and the Fonds
FQRNT of Qu\'ebec.  The  DRAO Synthesis
Telescope is operated as a national facility by the National Research 
Council of Canada. The CGPS is a Canadian project with
international partners and is supported by grants from NSERC.
Data from the CGPS
is publicly available through the facilities of the Canadian
Astronomy Data Centre (http://cadc.hia.nrc.ca) operated by the
Herzberg Institute of Astrophysics, NRC.

\end{document}